\documentstyle[12pt]{article}

\font\tenrm=cmr10

\begin{document}
\newcommand{\s}{\\ \vspace*{-4mm} }
\newcommand{\nn}{\noindent}
\newcommand{\non}{\nonumber}
\newcommand{\ee}{e^+ e^-}
\newcommand{\ra}{\rightarrow}
\newcommand{\lra}{\longrightarrow}
\newcommand{\beq}{\begin{eqnarray}}
\newcommand{\eeq}{\end{eqnarray}}
\newcommand{\tb}{{\rm tg} \beta}

\newcommand{\lsim}{\raisebox{-0.13cm}{~\shortstack{$<$ \\[-0.07cm] $\sim$}}~}
\newcommand{\gsim}{\raisebox{-0.13cm}{~\shortstack{$>$ \\[-0.07cm] $\sim$}}~}

\renewcommand{\thefootnote}{\fnsymbol{footnote} }

\begin{flushright}
\hfill PM/96--34\\ 
\hfill KA-TP-27-1996\\
\hfill December 1996\\
\end{flushright}

\vglue 0.7cm
\begin{center}{{\bf HIGGS PHENOMENOLOGY: A SHORT REVIEW}\\
\vglue 1.0cm
{\sc Abdelhak Djouadi}\\
\vglue 0.5cm 
{\it Laboratoire de Physique Math\'ematique et Th\'eorique\footnote{Present 
address; Unit\'e  UPRES--A 5032 associ\'ee au CNRS.} \\ 
Universit\'e de Montpellier II, F--34095 Montpellier Cedex 5, France.}\\

\vspace*{3mm}
{\it and} 
\vspace*{3mm}

{\it Institut f\"ur Theoretische Physik, Universit\"at Karlsruhe, \\
Kaiserstrasse 12, D--76128 Karlsruhe, Germany}\\

\vglue 1.5cm
{\sc Abstract}}
\end{center}
{\rightskip=3pc
 \leftskip=3pc
\tenrm\baselineskip=14pt

\noindent I briefly review the Higgs sector in the Standard Model and in
its minimal supersymmetric extension. After summarizing the properties of
the Higgs bosons, I will discuss the prospects for discovering these
particles at the present colliders LEP2 and Tevatron, and at the next 
generation colliders LHC and a high--energy $e^+e^-$ linear collider. The 
possibilities of studying the properties of the Higgs bosons will be then 
summarized. 

\vspace*{3cm}

\normalsize

\noindent Short write--up of lectures given at the {\em XXXVI Cracow 
School of Theoretical Physics}, Zakopane, Poland, June 1--10, 1996.}

\newpage

\setcounter{page}{2}
\clearpage
\def\thefootnote{\arabic{footnote}}
\setcounter{footnote}{0}

\subsection*{1. Introduction}

The search for Higgs particles is one of the main missions of present
and future high--energy colliders. The observation of this particle is
of utmost importance for the present understanding of the interactions
of the fundamental particles. Indeed, in order to accomodate the
well--established electromagnetic and weak interaction phenomena, the
existence of at least one isodoublet scalar field to generate fermion
and weak gauge bosons masses is required. The Standard Model (SM) makes
use of one isodoublet field: three Goldstone bosons among the four 
degrees of freedom are absorbed to build up the longitudinal components of 
the massive $W^\pm,Z$ gauge bosons; one degree of freedom is left over 
corresponding to a physical scalar particle, the Higgs boson \cite{1}. 
Despite of its numerous successes in explaining the present data, the 
Standard Model will not be completely tested before this particle has 
been experimentally observed and its fundamental properties studied. \s

In the Standard Model, the mass of the Higgs particle is a free
parameter. The only available information is the upper limit $M_H \gsim 
65$ GeV established at LEP1 \cite{2}, although the high--precision
electroweak data from LEP and SLC seem to indicate that its mass is
smaller than a few hundred GeV \cite{2}. However, interesting
theoretical constraints can be derived from assumptions on the energy
range within which the model is valid before perturbation theory breaks
down and new phenomena would emerge: \s

-- If the Higgs mass were larger than $\sim$ 1 TeV, the $W$ and $Z$ 
bosons would interact strongly with each other to ensure unitarity 
in their scattering at high energies. \s

-- The quartic Higgs self--coupling, which at the scale $M_H$ is
fixed by $M_H$ itself, grows logarithmically with the energy scale. If
$M_H$ is small, the energy cut--off $\Lambda$ at which the coupling
grows beyond any bound and new phenomena should occur, is large;
conversely, if $M_H$ is large, $\Lambda$ is small. The condition $M_H 
\lsim \Lambda$ sets an upper limit on the Higgs mass in the SM; lattice
analyses lead to an estimate of about 630 GeV for this limit.
Furthermore, top quark loops tend to drive the coupling to negative
values for which the vacuum is no more stable. Therefore, requiring the
SM to be extended to the GUT scale, $\Lambda_{\rm GUT} \sim 10^{15}$
GeV, and including the effect of top quark loops on the running
coupling, the Higgs boson mass should roughly lie in the range between 
100 and 200 GeV; see Ref.~\cite{3} for a recent discussion. \s

However, there are two problems that one has to face when trying to
extend the SM to $\Lambda_{\rm GUT}$. The first one is the so--called
hierarchy or naturalness problem: the Higgs boson tends to acquire a
mass of the order of these large scales [the radiative corrections to
$M_H$ are quadratically divergent]; the second problem is that the
simplest GUTs predict a value for $\sin^2\theta_W$ that is incompatible
with the measured value $\sin^2\theta_W \simeq~0.23$. Low energy 
supersymmetry solves these two problems at once: supersymmetric particle 
loops cancel exactly the quadratic divergences and contribute to the 
running of the gauge coupling constants, correcting the small discrepancy 
to the observed value of $\sin^2\theta_W$; see Ref.~\cite{3} for a recent
review. \s

The minimal supersymmetric extension of the Standard Model
(MSSM) requires the existence of two isodoublets of Higgs
fields, to cancel anomalies and to give mass separately to up and
down--type fermions \cite{1}. Three neutral, $h/H$(CP=+), $A$(CP=--) 
and a pair
of charged scalar particles, $H^\pm$, are introduced by this extension
of the Higgs sector. Besides the four masses, two additional parameters
define the properties of these particles: a mixing angle $\alpha$ in the
neutral CP--even sector and the ratio of the two vacuum expectation
values $\tb$, which from GUT restrictions is assumed in the range $1 \lsim
\tb \lsim m_t/m_b$ with the lower and upper ranges favored by Yukawa
coupling unification. \s

Supersymmetry leads to several relations among these
parameters and only two of them are in fact independent. These relations
impose a strong hierarchical structure on the mass spectrum, $M_h<M_Z,
M_A<M_H$ and $M_W<M_{H^\pm}$, which however is broken by radiative
corrections if the top quark mass is large \cite{4}. For instance, the
upper bound on the mass of the lightest Higgs boson $h$ is shifted from
the tree level value $M_Z$ to $\sim 130$ GeV. The masses of the 
heavy neutral and charged Higgs particles can be expected, with a high
probability, in the range of the electroweak symmetry breaking scale. \s

Some of these features are not specific to the minimal extension and are
expected to be realized also in more general SUSY models. For instance,
a light Higgs boson with a mass below ${\cal O}$(200 GeV) is quite
generally predicted by SUSY theories \cite{5}. \s

The search for these Higgs particles will be the major goal of the next
generation of colliders. In the following, after summarizing the properties
of the Higgs bosons, I will briefly discuss the discovery potential of the 
present colliders LEP2 \cite{6} and Tevatron \cite{7} as well as the pp 
collider LHC [8--10] with a c.m. energy of $\sim 14$ TeV and a future 
${\rm \ee}$ linear collider [11--13] with a c.m. energy in the range 
of 300 to 500 GeV. The case of future muon colliders is discussed in 
Ref.~\cite{13}. More detailed discussions and a complete set of references 
can be found in Refs.~[6--14]. 

\subsection*{2. Couplings and Decay Modes} 

\subsubsection*{2.1 Standard Higgs boson}

In the SM, the profile of the Higgs particle is uniquely determined once
$M_H$ is fixed. The decay width, the branching ratios and the production
cross sections are given by the strength of the Yukawa couplings to
fermions and gauge bosons, the scale of which is set by the masses of
these particles. To discuss the Higgs decay modes \cite{1,14}, it is 
convenient to divide the Higgs mass into two ranges: the ``low mass" range 
$M_H \lsim 130$ GeV and the ``high mass" range $M_H \gsim 130$ GeV. \s

In the ``low mass" range, the Higgs boson decays into a large variety of
channels. The main decay mode is by far the decay into $b\bar{b}$ pairs
with a branching ratio of $\sim 90\%$ followed by the decays into
$c\bar{c}$ and $\tau^+\tau^-$ pairs with a branching ratio of $\sim
5\%$. Also of significance, the top--loop mediated Higgs decay into
gluons, which for $M_H$ around 120 GeV occurs at the level of $\sim
5\%$. The top and $W$--loop mediated $\gamma\gamma$ and $Z \gamma$ decay
modes are very rare the branching ratios being of ${\cal O }(10^{-3})$;
however these decays lead to clear signals and are interesting being
sensitive to new heavy particles. \s

In the ``high mass" range, the Higgs bosons decay into $WW$ and $ZZ$
pairs, with one of the gauge bosons being virtual below the threshold.
Above the $ZZ$ threshold, the Higgs boson decays almost exclusively into
these channels with a branching ratio of 2/3 for $WW$ and 1/3 for $ZZ$.
The opening of the $t\bar{t}$ channel does not alter significantly this
pattern, since for large Higgs masses, the $t\bar{t}$ decay width rises
only linearly with $M_H$ while the decay widths to $W$ and $Z$ bosons
grow with $M_H^3$. \s

In the low mass range, the Higgs boson is very narrow $\Gamma_H<10$ MeV,
but the width becomes rapidly wider for masses larger than 130 GeV,
reaching 1 GeV at the $ZZ$ threshold; the Higgs decay width cannot be
measured directly in the mass range below 250 GeV. For large masses, 
$M_H \gsim 500$ GeV, the Higgs boson becomes obese: its decay width becomes 
comparable to its mass. \s

\subsubsection*{2.2 MSSM Higgs bosons}

In the MSSM, one usually chooses the basic parameters to be the mass of
the pseudoscalar Higgs boson $M_A$ and $\tb$. Once these two parameters
are specified, all other masses and the angle $\alpha$ can be derived at
the tree--level. These relations are however affected by radiative
corrections, the leading part of which grows as $m_t^4$ and
logarithmically with the common squark mass. Subleading corrections will
introduce the soft--SUSY breaking trilinear coupling $A_{t,b}$ and
Higgs--higgsino mass parameter $\mu$. \s

The couplings of the various neutral Higgs bosons [collectively denoted
by $\Phi$] to fermions and gauge bosons will in general strongly depend
on the angles $\alpha$ and $\beta$; normalized to the SM Higgs
couplings, they are given by 

\smallskip

\begin{center}
\begin{tabular}{|c|c|c|c|c|} \hline
$\ \ \ \Phi \ \ \ $ &$ g_{\Phi \bar{u}u} $	& $ g_{\Phi \bar{d} d} $ &
$g_{ \Phi VV} $ \\ \hline
$h$  & \ $\; \cos\alpha/\sin\beta	\; $ \ & \ $ \;	-\sin\alpha/
\cos\beta \; $ \ & \ $ \; \sin(\beta-\alpha) \;	$ \ \\
 $H$  & \	$\; \sin\alpha/\sin\beta \; $ \	& \ $ \; \cos\alpha/
\cos\beta \; $ \ & \ $ \; \cos(\beta-\alpha) \;	$ \ \\
$A$  & \ $\; 1/ \tb \; $\ & \ $	\; \tb \; $ \	& \ $ \; 0 \; $	\ \\ \hline
\end{tabular}
\end{center}
\vspace{0.3cm}

The pseudoscalar has no tree level couplings to gauge bosons, and its
couplings to down (up) type fermions are (inversely) proportional to
$\tb$. It is also the case for the couplings of the charged Higgs
particle to fermions which are a mixture of scalar and pseudoscalar
currents and depend only on $\tb$. For the CP--even Higgs bosons, the
couplings to down (up) type fermions are enhanced (suppressed) compared
to the SM Higgs couplings [$\tb \gsim 1$]. If the pseudoscalar mass is large,
the $h$ boson reaches its upper limit [which depends on the value of
$\tb$] and its couplings to fermions and gauge bosons are SM like; the
CP--even and charged Higgs bosons $H$ and $H^\pm$ will be degenerate
with $A$. In this decoupling limit, it is very difficult to distinguish the
Higgs sector of the MSSM from the one of the SM. \s 

Since its mass is smaller than $\sim$ 130 GeV, the lightest Higgs boson
will decay mainly into fermion pairs, mostly $b\bar{b}$ and $\tau^+ \tau^-$
pairs since the couplings of these particles are enhanced for $\tb>1$;
the decays into $c\bar{c}$ as well as the $gg$ decay are in general
strongly suppressed especially for large values of $\tb$. In the
decoupling limit, the $h$ branching ratios become SM--like and for
masses close to 130 GeV, the $h \ra WW^*$ becomes of some relevance. The
two--photon decay mode of the $h$ is suppressed in general compared to
the SM. \s

The decay pattern of the heavier MSSM Higgs bosons \cite{1,14} depend 
strongly on the value of $\tb$. For large $\tb$ values, the pattern is 
quite simple: the neutral Higgs boson $H$ and $A$ will mainly decay into 
$b \bar{b}$ and $\tau^+ \tau^-$ pairs with branching ratios close to 90\% 
and 10\% respectively, and the charged Higgs boson will decay into $\tau 
\nu_\tau$ or $tb$ pairs, depending on whether it is lighter or heavier than 
the top quark. \s

For small values of $\tb$ the situation is simple only above the $2m_t$
[$m_t+m_b$] threshold for neutral [charged] Higgs bosons: the decay
channels $H,A \rightarrow t\bar{t}$ and $H^+ \ra t\bar{b}$ are then 
dominating. Below the top threshold, the $H$ boson mainly decays into two 
light Higgs bosons $H \ra hh$, while the pseudoscalar decays into the lightest
Higgs boson $h$ and a $Z$ boson. The decays into $b\bar{b}$ and {\it a 
fortiori} the decays into the lighter fermions are rare in general; for
$b\bar{b}$ they are important
only for small $A,H$ masses when the channel $A \ra Zh$ is closed or the
trilinear $Hhh$ coupling is small. The decays into $WW/ZZ$ pairs are
suppressed for $H$ and due to CP invariance are absent in the case of $A$. 
For the charged Higgs boson, the decays $H^\pm \ra hW^\pm$ [and if $A$ is
light, the decays $H^\pm \ra A W^\pm$] are also of significance below the 
$tb$ threshold. \s

The branching ratios of the $\gamma \gamma$ and $Z \gamma$ decays are
smaller than in the SM; this is due to the fact that the $b\bar{b}$ decays 
are enhanced for $\tb \gsim 1$ and the dominant $W$--loop contribution 
is suppressed (absent) in the case of the CP--even (odd) Higgs bosons. \s

Other possible channels are the decays into SUSY particles. Indeed the
Higgs decays into charginos and neutralinos could be very important
since some of these particles are expected to have masses of ${\cal
O}(M_Z$). For small values of $\tb$ and below the $t\bar{t}$ threshold,
these decays become in fact dominant. The decays into squarks, and in 
particular top squarks can also be very important for small $\tb$ 
since the Higgs couplings to top squark is very strong [proportional to 
$m_t^2$]; in fact, when they are kinematically allowed, these channels 
are the dominant ones. Decays of the Higgs bosons into sleptons when 
kinematically allowed are marginal. \s

Adding up the various decay modes, the width of all five Higgs bosons
are relatively narrow compared to the SM case: for small masses they are
below a few GeV, while for masses $\sim 1$ TeV they can reach values
of order a few ten GeV if $\tb$ is extremely large. This has to be contrasted
with the SM where the width becomes comparable to its masses in the TeV
range: in the MSSM, the decay into massive gauge boson pairs is
either absent or strongly suppressed and the widths increase only 
linearly with the Higgs masses. 

\subsection*{3. Higgs searches at Present colliders}

\subsubsection*{3.1 Searches at LEP}

The most comprehensive search of Higgs bosons done so far was undertaken 
by the LEP experiments \cite{2}. In the SM, the main production process is the 
so--called Bjorken or bremsstrahlung process, where the $Z$ resonance
emits a Higgs boson, turns virtual and decays into two massless fermions
\beq
(a) \hspace*{1cm} Z \ra Z^* H \ra H f\bar{f} \nonumber
\eeq
Although the virtuality of the $Z$ boson is penalizing since the cross
section is suppressed by a power of the electroweak coupling, the large
number of $Z$ bosons collected at LEP1 allows to have a sizeable rate
for not too heavy Higgs bosons. From the negative search of such events,
a lower bound of $M_H \gsim 65$ GeV has been set. Note that even
almost massless Higgs bosons have been ruled out using this process:
indeed, even in this case, the Higgs particle will carry momentum and
will alter the kinematics of the visible final $Z^* \ra f\bar{f}$ state. \s 

The LEP collaborations have also searched for the MSSM lightest CP--even 
Higgs boson $h$ and for a light pseudoscalar particle $A$. The $h$ boson 
can be produced in the Bjorken process as in the case of the SM Higgs, 
but here the cross section is suppressed by the $hZZ$ coupling squared 
$\sin^2 (\beta-\alpha)$. The $h$ boson can also be produced in association 
with the pseudoscalar Higgs boson $A$ 
\beq
(b) \hspace*{1cm} Z \ra h A \nonumber
\eeq
The cross section is suppressed by a factor $\cos^2(\beta-\alpha)$ and
therefore this process is complementary to the bremsstrahlung process. For
$A$ masses below $M_A \lsim M_Z/2$, the sum of the cross sections of the 
two production channels is always large; a bound of $M_h \sim M_A \gsim 45$ 
GeV has been set on the two particles. \s

At LEP2 \cite{6}, with a center of mass energy above the $2M_W$ threshold, the
SM Higgs boson will be searched for using the same process $(a)$ with
the difference that now the final $Z$ boson is on--shell. The process is thus 
at lowest order in the electroweak coupling and gives a decent cross 
section, although we are no more on the $Z$ resonance. The Higgs bosons 
will mainly decay into $b\bar{b}$ final states, 
requiring efficient means to tag the $b$--quark jets. The backgrounds are 
rather small, except for the process $\ee \ra ZZ \ra b\bar{b}Z$ for Higgs 
masses close to $M_Z$. Depending on the final energy which will be reached 
at LEP2, $\sqrt{s}=175$ or $192$ GeV, Higgs masses close to 80 and 90 GeV, 
respectively, can be probed with an integrated luminosity of $\int {\cal L} 
=150$ pb$^{-1}$ \cite{6}. \s

Similarly to the LEP1 case, the MSSM Higgs bosons can be searched for in
the bremsstrahlung and the associated production processes $\ee \ra hZ$
and $\ee \ra Ah$ with the $Z$ boson being on--shell for the first channel.
The $h$ and $A$ bosons will mainly decay into $b\bar{b}$ final states. 
Again the background events are rather small, and using the complementarity 
of the two processes, the range $M_h \lsim 80$ or $90$ GeV can be probed
with a luminosity $\int {\cal L} \sim 150$ pb$^{-1}$ at c.m. energies
$\sqrt{s}=175$ or $192$ respectively \cite{6}. This means that the entire 
range 
for the $h$ mass, $M_h \lsim 80$ GeV, for small values of $\tb \lsim 1.5$ 
[which are favored by the requirement of $b$--$\tau$ Yukawa coupling
unification] can be probed at LEP2. If the Higgs boson has a mass in the
range $90 \lsim M_h \lsim 130$ GeV [for larger $\tb$ values], its production
will have to await for the next generation of colliders.  
  
\subsubsection*{3.2 Searches at the Tevatron}

Currently, the Fermilab Tevatron collider is operating at a c.m. 
energy $\sqrt{s}=1.8$ TeV with a luminosity ${\cal L} \sim 10^{31}$ 
cm$^{-2}$s$^{-1}$. With the main injector, which is expected to begin 
operation in a few years, the luminosity will be increase to ${\cal L} \sim 
2.10^{32}$ cm$^{-2}$s$^{-1}$ and the c.m. energy to $\sqrt{s}=2$ TeV. An 
increase of the luminosity to the level of ${\cal L} \sim 10^{33}$ 
cm$^{-2}$s$^{-1}$ [the so--called TEV33 option] is also currently 
discussed \cite{7}. \s

The most promising production mechanism of the SM Higgs boson at the 
Tevatron collider is the $W H$ process, with the Higgs boson decaying 
into $b\bar{b}$ [or $\tau^+ \tau^-$] pairs
\beq
(c) \hspace*{1cm} qq \ra W H \ra W b \bar{b} \nonumber
\eeq
For Higgs masses $M_H \sim 100$ GeV, the cross section is of the order 
of a few tenths of a picobarn. The related production process $q \bar{q} 
\ra ZH \ra Z b\bar{b}$ [which is the equivalent of the Bjorken process 
in pp collisions] has a smaller cross section, a result of the small 
neutral current couplings compared to charged current  couplings. \s

The main irreducible backgrounds will consist of $W b \bar{b}$ and $WZ 
\ra Wb\bar{b}$ for $M_H \sim M_Z$, as well as $t\bar{t}$ production for
$M_H \gsim 100$ GeV. These backgrounds have cross sections which are 
of the same order as the signal cross section; an important issue will be 
the $b\bar{b}$ invariant mass resolution which needs to be measured with 
a very good accuracy. The 
Higgs search at the Tevatron with a luminosity of $\sim 2$ fb$^{-1}$ will 
probably be limited to $ M_H \lsim M_Z$ \cite{7}, a mass region which will 
be already covered at LEP2. To probe Higgs masses larger than $M_Z$, a higher 
luminosity will be required, and the TEV33 option will be mandatory. A 
rather detailed analysis for TEV33 with an integrated luminosity  of 
$\int {\cal L} \sim 30$ fb$^{-1}$, concluded that Higgs masses up to $M_H 
\sim 120$ GeV could possibly be reached \cite{7}. The processes $WH, ZH$ with 
$H \ra \tau^+ \tau^-$ and $W,Z \ra2$ jets will not significantly change
this picture. \s

In the MSSM, the only useful production mechanism will also be the $Wh$ 
strahlung process with $h \ra b\bar{b}$, since the associated production 
mechanism $q\bar{q} \ra Ah$ would have a too small cross section as in the 
case of the $q\bar{q} \ra Zh$ process. However, the $Wh$ production cross 
section will be suppressed by a factor $\sin^2(\beta-\alpha)$ compared to 
the SM Higgs boson case, except if the $h$ mass is maximal for a given value 
of $\tb$ for which $\sin^2(\beta-\alpha)=1$. In this case, $h$ is almost SM 
like and the discovery reach of the Tevatron will be the same as previously 
discussed.

\subsection*{4. Production at LHC} 

\subsubsection*{4.1 SM Higgs Boson}

The main production mechanisms of neutral Higgs bosons at hadron colliders
are the following processes \cite{8,9}
\begin{eqnarray}
\begin{array}{lccl}
(a) & \ \ {\rm gluon-gluon~fusion} & \ \ gg  \ \ \ra & H \nonumber \\
(b) & \ \ WW/ZZ~{\rm fusion}       & \ \ VV \  \ra &  H \nonumber \\
(c) & \ \ {\rm association~with}~W/Z & \ \ q\bar{q} \ \ \ra & V + H \nonumber
\\
(d) & \ \ {\rm association~with~}\bar{t}t & gg,q\bar{q}\ra & t\bar{t}+H
\nonumber
\end{array}
\end{eqnarray}

In the interesting mass range, $100 \lsim M_H \lsim 200$ GeV, the dominant 
production process of the SM Higgs boson is the gluon--gluon fusion mechanism 
[in fact, it is the case of the entire Higgs mass range] for which the cross 
section is of order a few tens of pb. 
It is followed by the $WW/ZZ$ fusion processes [especially for large $M_H$] 
with a cross section of a few pb; the cross sections of the associated
production with $W/Z$ or $t\bar{t}$ are an order of magnitude smaller. Note
that for a luminosity of ${\cal L}=10^{33} (10^{34})$~cm$^{-2}$s$^{-1}$, 
$\sigma=$~1 pb would correspond to $10^{4}(10^{5})$ events per year. \s

Besides the errors due to the relatively poor knowledge of the gluon 
distribution at small $x$, the lowest order cross sections are affected 
by large uncertainties due to higher order corrections. Including the 
next to leading QCD corrections, the total cross sections can be defined 
properly: the scale at which one defines the strong coupling constant 
is fixed and the [generally non--negligible] corrections are taken into 
account. The ``K--factors" for $WH/ZH$ production [which can be inferred 
from the Drell--Yan $W/Z$ production] and the $VV$ fusion mechanisms 
are small, increasing the total cross sections by $\sim$ 20 and 10\% 
respectively;  the QCD corrections to the associated $t\bar{t}H$ production 
are still not known. The [two--loop] QCD corrections to the main mechanism, 
$gg \ra H$, have been computed \cite{15} and found to be rather large 
since they increase the cross sections by a factor $\simeq 1.8$ at LHC [there 
is, however, an uncertainty of $\sim 20\%$ due to the arbitrariness of the 
choice of the renormalization and factorization scales and also of the 
parton densities]. \s

The signals which are best suited to identify the produced Higgs particles
at the LHC have been studied in great detail in Refs.~[8,9]. I briefly
summarize here the main conclusions of these studies.\s

For Higgs bosons in the ``high mass" region, $M_H \gsim 130$~GeV, the signal
consists of the so--called ``gold--plated" events $H \ra Z Z^{(*)} \ra 4l^\pm$
with $l=e,\mu$. The backgrounds [mostly $pp \ra ZZ^{(*)}, Z \gamma^*$ for the 
irreducible background and $t \bar{t} \ra WWb \bar{b}$ and $Zb \bar{b}$ for the 
reducible one] are relatively small. One can probe Higgs masses up to ${\cal 
O}$(700~GeV) with a luminosity $\int {\cal L}= 100 $~fb$^{-1}$ at LHC. 
The $H \ra WW^{(*)}$ decay channel is more difficult to use because of the 
large background from $t\bar{t}$ pair production; the $H \ra t\bar{t}$
signal is swamped by the irreducible background from $gg\ra t\bar{t}$. 
For $M_H \gsim 700$ GeV
[where the Higgs boson total decay width becomes very large], the search 
strategies become more complicated; see Ref.~\cite{8}. \s

For the ``low mass" range, the situation is more complicated. The branching
ratio for $H\ra ZZ^*$ becomes too small and due to the huge QCD jet background,
the dominant mode $H\ra b\bar{b}$ is practically 
useless; one has then to rely on the rare
$\gamma \gamma$ decay mode with a branching ratio of ${\cal O}(10^{-3})$. At
LHC with a luminosity of $\int {\cal L}= 100$~fb$^{-1}$, the cross section
times the branching ratio leads to ${\cal O}(10^{3})$ events but one has to
fight against formidable backgrounds. Jets faking photons need a rejection
factor larger than $10^{8}$ to be reduced to the level of the physical
background $q\bar{q}, gg \ra \gamma \gamma$ which is still very large. However,
if very good geometric resolution and stringent isolation criteria, combined
with excellent electromagnetic energy resolution to detect the narrow $\gamma
\gamma$ peak of the Higgs boson are available [one also needs a high
luminosity $ {\cal L} \simeq 10^{34}$~cm$^{-2}$s$^{-1}$], this channel,
although very difficult, is feasible: for $\int {\cal L}=100$ fb$^{-1}$,
ATLAS claims a sensitivity for $110$ GeV $ \lsim M_H \lsim 140$ GeV and 
requires five times more luminosity to reach down masses $M_H \sim 80$ GeV; 
CMS [which benefits from a good electromagnetic calorimeter] claims a 
coverage  $85$ GeV $\lsim M_H \lsim 150$ GeV for 100 fb$^{-1}$. The low end
of the mass range is the most challenging due to the small branching ratio
and the larger backgrounds. \s

Complementary production channels would be the $pp \ra WH, t\bar{t}H 
\ra \gamma \gamma l \nu$ processes for which the backgrounds are much smaller
since one requires an additional lepton. However the signal cross sections 
are very small too, making these processes also difficult. The processes $pp 
\rightarrow WH$ and $t\bar{t}H$ with $H \ra b\bar{b}$ 
seem also promising provided that very good micro--vertexing to tag the 
$b$--quarks can be achieved. 

\subsubsection*{4.2 MSSM Higgs Bosons}

In the MSSM, the situation is more difficult than in the SM. The 
production mechanisms of the neutral SUSY Higgs bosons are practically 
the same as those of the SM Higgs; one only has to take the $b$ quark 
[whose couplings are strongly enhanced for large $\tb$ values] contributions 
into account in the $gg \ra$ Higgs process [and also the extra contributions 
from squarks loops, which however decouple for high squark masses] and 
for $\tb\gg1$, the $q\bar{q}\ra b\bar{b}+A/H$ processes become the dominant 
production mechanisms for the $H$ and $A$ bosons. The cross sections are 
the same as in the case of the SM, modulo mixing angle suppression/enhancement 
factors.\s

The various signals for the SUSY Higgs bosons can be summarized as follows: \s

i) Since the lightest Higgs boson mass is always smaller than $\sim 130 $
GeV, the $ZZ$ signal cannot be used. Furthermore, the $hWW(h\bar{b}b)$ coupling
is suppressed (enhanced) leading to a smaller $\gamma \gamma$ branching ratio
than in the SM [additional contributions from chargino and sfermion loops can
also alter the decay width] making the search more difficult. If $M_h$ is
close to its maximum value, $h$ has SM like couplings and the situation is
similar to the SM case with $M_H \sim $ 80--130 GeV. \s

ii) Since $A$ has no tree--level couplings to gauge bosons and since the 
couplings of the heavy CP--even $H$ are strongly suppressed, the gold--plated 
$ZZ$ signal is lost [for $H$ it survives only for small $\tb$ values, provided 
that $M_H<2m_t$]. In addition, the $A, H \ra \gamma \gamma$ signals cannot be 
used since the branching ratios are suppressed.  One
has then to rely on the $A,H  \ra \tau^+ \tau^-$ channels for large $\tb$ 
values; this mode, which is hopeless for the SM Higgs, seems to be feasible in 
this case. For small $\tb$ and below the $t\bar{t}$ threshold, the cascade 
decays $H \ra hh \ra b \bar{b}b\bar{b}$ and $A \ra hZ \ra Zb \bar{b}$ have 
large branching ratios and can be employed if efficient $b$--tagging is 
available at LHC. The decays $H/A \ra t\bar{t}$ are challenging to detect. \s  

iii) Charged Higgs particles, if lighter than the top quark, can be
accessible in top decays $t \ra H^+b$. This results in a surplus of $\tau$
lepton final states [the main decay mode is $H^-\ra \tau \nu_\tau$] over
$\mu,e$ final states, an apparent breaking of $\tau \ vs. \ e,\mu$
universality. At LHC, $H^\pm$ masses up to $\sim 140$ GeV can be probed for 
$m_t \sim 175$ GeV. Additional improvements in some areas of the parameter
space can be made by considering the process $gg \ra tb H^\pm \ra t\bar{t}b
\bar{b}$ with efficient $b$--tagging. \s

iv) All the previous discussion assumes of course that the decays into 
supersymmetric particles are kinematically inaccessible. This seems to 
be very unlikely, since at least the decays of the heavy Higgs particles
$H,A$ and $H^\pm$ into charginos and neutralinos should be possible. If
this scenario is realised, the discovery of these Higgs particles will be
even more challenging. If the lightest neutralinos are lighter than 
$M_h/2$, then the discovery of the lightest Higgs [and also of the heavier
ones] will be practically impossible at LHC. \s

Thus, the search for SUSY Higgs bosons is more difficult than the search
for the SM Higgs, especially if the decays into SUSY particles are possible. 
If SUSY decays are kinematically not allowed, detailed 
analyses of the ATLAS and CMS collaborations concluded that at least
one Higgs boson will be discovered in the entire MSSM parameter space,
after a few years of running with a luminosity of $10^{34}$cm$^{-2}$s$^{-1}$.
 
\subsection*{5. Production at e$^+$e$^-$ Colliders}

\subsubsection*{5.1 SM Higgs boson}

At $\ee$ linear colliders operating in the 500 GeV energy range  the
main production mechanisms for SM Higgs particles are \cite{11,12}
\begin{eqnarray}
\begin{array}{lccl}
(a)  & \ \ {\rm bremsstrahlung \ process} & \ \ \ee & \ra (Z) \ra Z+H \non \\
(b)  & \ \ WW \ {\rm fusion \ process} & \ \ \ee & \ra \bar{\nu} \ \nu \
(WW) \ra \bar{\nu} \ \nu \ + H \non \\
(c)  & \ \ ZZ \ {\rm fusion \ process} & \ \ \ee & \ra e^+ e^- (ZZ) \ra
e^+ e^- + H \non \\
(d)  & \ \ {\rm radiation~off~tops} & \ \ \ee & \ra (\gamma,Z) \ra
t \bar{t}+H \non
\end{array}
\end{eqnarray}

The Higgs--strahlung cross section scales as $1/s$ and therefore dominates 
at low energies while the $WW$ fusion mechanism  has a cross section 
which rises like $\log(s/M_H^2)$ and dominates at high energies. 
At $\sqrt{s} \sim 500$ GeV, the two processes have approximately 
the same cross sections for the interesting range 100 GeV $\lsim M_H 
\lsim$ 200 GeV. With an in integrated luminosity $\int {\cal L}
\sim 50$ fb$^{-1}$, approximately 2000 events per year can be collected
in each channel; a sample which is more than enough to discover the
Higgs boson and to study it in detail. The $ZZ$ fusion mechanism $(c)$
and the associated production with top quarks $(d)$ have much smaller 
cross sections. But these processes will be very useful when it comes 
to study the Higgs properties as will be discussed later. \s

In the Higgs-strahlung process, the recoiling $Z$ boson
[which can be tagged through its clean $\mu^+ \mu^-$ decay] is
mono--energetic and the Higgs mass can be derived from the energy of the
$Z$ if the initial $e^+$ and $e^-$ beam energies are sharp
[beamstrahlung, which smears out the c.m. energy should thus be suppressed as
strongly as possible, and this is already the case for machine designs
such as TESLA]. Therefore, it will be easy to separate the signal from
the backgrounds. For low Higgs masses, $M_H \lsim 130$ GeV, the
main background will be $\ee \ra ZZ$. The cross section is large, but it
can be reduced by cutting out the forward and backward directions 
[the process is
mediated by $t$--channel $e$ exchange] and by selecting $b\bar{b}$
final states by means of $\mu$--vertex detectors [while the Higgs decays
almost exclusively into $b\bar{b}$ in this mass range, BR$(Z \ra
b\bar{b}$) is small, $\sim 15\%$]. The background from single $Z$
production, $\ee \ra Zq\bar{q}$, is small and can be further reduced by
flavor tagging. In the mass range where the decay $H \ra WW^*$ is
dominant, the main background is triple gauge boson production and is
suppressed by two powers of the electroweak coupling. \s

The $WW$ fusion mechanism offers a complementary 
production channel. For small $M_H$, the main backgrounds are single $W$ 
production, $\ee \ra e^\pm W^\mp \nu$ $[W \ra q\bar{q}$ and the $e^\pm$ 
escape detection] and $WW$ fusion into a $Z$ boson, $\ee \ra \nu 
\bar{\nu}Z$, which have cross sections 60 and 3 times larger than the
signal, respectively. Cuts on the rapidity spread, the energy and
momentum distribution of the two jets in the final state [as well as 
flavor tagging for small $M_H$] will suppress these background events. \s

It has been shown in detailed simulations, that just a few fb$^{-1}$ 
of integrated luminosity are needed to obtain a 5$\sigma$ signal for
a Higgs boson with a mass $M_H \lsim 140$ GeV at a 500 GeV collider 
[in fact, in this case, it is better to go to lower energies where the 
cross section is larger], even if it decays invisibly [as it could 
happen in SUSY models for instance]. Higgs bosons with masses 
up to $M_H \sim 350$ GeV can be discovered at the 5$\sigma$ level, in both 
the strahlung and fusion processes at an energy of 500 GeV and with a
luminosity of 50 fb$^{-1}$. For even higher masses, one needs to 
increase the c.m. energy of the collider, and as a rule of thumb, Higgs 
masses up to $\sim 70\%$ of the total energy of the collider can be 
probed. This means than a $\sim 1$ TeV collider will be needed to probe the 
entire Higgs mass range in the SM. 

\subsubsection*{5.2 MSSM Higgs bosons}

An even stronger case for $\ee$ colliders in the 300--500 GeV energy range
is made by the MSSM. In $\ee$ collisions, besides the usual
bremsstrahlung and fusion processes for $h$ and $H$ production, the neutral
Higgs particles can also be produced pairwise: $\ee \ra A + h/H$.
The cross sections for the bremsstrahlung and the pair production as
well as the cross sections for the production of $h$ and $H$ are mutually
complementary, coming either with a coefficient $\sin^2(\beta-
\alpha)$ or $\cos^2(\beta -\alpha)$. The cross section for $hZ$ production is
large for large values of $M_h$, being of ${\cal O}(50$ fb); by contrast, the
cross section for $HZ$ is large for light $h$ [implying small $M_H$].  In major
parts of the parameter space, the signals consist of a $Z$ boson and a
$b\bar{b}$ or a $\tau^+ \tau^-$ pair, which is easy to separate from the main
background, $\ee \ra ZZ$ [for $M_h \simeq M_Z$, efficient $b$ detection is
needed]. For the associated production, the situation is opposite: the cross
section for $Ah$ is large for light $h$ whereas $AH$ production is preferred in
the complementary region.  The signals consists mostly of four $b$ quarks in
the final state, requiring efficient $b\bar{b}$ quark tagging; mass constraints
help to eliminate the QCD jets and $ZZ$ backgrounds. The CP--even Higgs
particles can also be searched for in the $WW$ and $ZZ$ fusion mechanisms. \s

In $\ee$ collisions, charged Higgs bosons can be produced pairwise, $\ee \ra
H^+H^-$ through $\gamma,Z$ exchange. The cross section depends only on the
charged Higgs mass; it is large up to $M_{H^\pm} \sim 230$~GeV. Charged Higgs
bosons can also be created in laser--photon collisions, $\gamma \gamma \ra
H^+H^-$ but due to the reduced energy, only smaller masses than previously can
be probed. The cross section, however, is enhanced in the low mass range.
Finally, charged Higgs bosons can be produced in top decays as discussed above
in the case of proton colliders. In the range $ 1 < \tb < m_t/m_b$, the $t \ra
H^+b$ branching ratio varies between $ \sim 2 \%$ and $20 \%$ and since the
cross section for $t\bar{t}$ production is ${\cal O}(0.5$~pb) at $\sqrt{s}=500$
GeV, this corresponds to 200 and 2000 charged Higgs bosons at a luminosity
$\int {\cal L} =$ 10 fb$^{-1}$. \s

The preceding discussion on the MSSM Higgs sector in $\ee$ linear colliders
can be summarized in the following points \cite{11,12}: \s

i) The Higgs boson $h$ can be detected in the entire range of the MSSM
parameter space, either through the bremsstrahlung process or through pair
production. In fact, this conclusion holds true even at a c.m.
energy of 300 GeV. \s

ii) There is a substantial area of the ($ M_h,\tb$) parameter space where
{\it all} SUSY Higgs bosons can be discovered at a 500 GeV collider.
This is possible if the $H,A$ and $H^{\pm}$ masses are less than $\sim 230$
GeV. For large Higgs masses, one simply has to increase the c.m. energy. \s
 
iii) Even if SUSY Higgs decays are allowed to occur, the Higgs particles
can be easily detected in $\ee$ collisions. Using the missing mass technique,
the CP--even $h$ and $H$ bosons can be detected in the bremsstrahlung process 
even if they decay invisibly. The pseudoscalar and the charged Higgs bosons
can also be detected by looking at mixtures of SUSY and standard
decays which occur at observable rates. \s 

vi) In some parts of the MSSM parameter space, the lightest Higgs $h$ can
be detected, but it cannot be distinguished from the SM Higgs boson. In this
case, Higgs production in $\gamma \gamma$ fusion [which receives extra
contributions from SUSY particle loops] can be helpful. 

\subsection*{6. Study of Higgs properties}

Once the Higgs boson is found it will be of great importance to explore 
all its fundamental properties. This can be done at great details especially 
in the clean environment of $\ee$ linear colliders: the Higgs mass, the
spin and parity quantum numbers and the couplings to fermions and
gauge bosons can measured. In the following we will summarize these
features in the case of the SM Higgs boson; some of this discussion
can be of course extended to the case of the lightest MSSM Higgs particle.

\subsubsection*{6.1 Studies at $\ee$ Colliders $[11,12]$}

In the Higgs--strahlung process with the $Z$ decaying into
visible particles, the mass resolution achieved with kinematical
constraints is close to 5 GeV, and a precision of about $\pm 200$ MeV
can be obtained on the Higgs mass with $\int {\cal L}=10$
fb$^{-1}$ if the effects of beamstrahlung are small. For masses below 
250 GeV, the Higgs boson is extremely narrow and its
width cannot be resolved experimentally; only for higher masses [or at
$\mu^+ \mu^-$ colliders, see \cite{13} e.g.] $\Gamma_H$ can be 
measured directly. \s

The angular distribution of the $Z/H$ in the Higgs--strahlung 
process is sensitive to the spin--zero of the Higgs particle: 
at high--energies the $Z$ is
longitudinally polarized and the distribution follows the $\sim
\sin^2\theta$ law which unambiguously characterizes the production of a
$J^P=0^+$ particle. The spin--parity quantum numbers of the Higgs
bosons can also be checked experimentally by looking at correlations in
the production $\ee \ra HZ \ra$ 4--fermions or decay $H \ra WW^* \ra$
4--fermion processes, as well as in the more difficult
channel $H \ra \tau^+ \tau^-$ for $M_H \lsim 140$ GeV. An unambiguous test
of the CP nature of the Higgs bosons can be made in the process $\ee \ra
tt \bar{H}$ or at laser photon colliders in the loop 
induced process $\gamma \gamma \ra H$. \s

The masses of the fermions are generated through the Higgs
mechanism and the Higgs couplings to these particles are proportional to
their masses. This fundamental prediction has to be verified
experimentally. The Higgs couplings to $ZZ/WW$ bosons can be
directly determined by measuring the production cross sections in the
bremsstrahlung and the fusion processes. In the $\ee \ra H\mu^+\mu^-$
process, the total cross section can be measured with a precision of
less than 10\% with 50 fb$^{-1}$. \s

The Higgs couplings to light fermions are harder to measure,
except if $M_H \lsim 140$ GeV. The Higgs branching ratios to $b\bar{b}$,
$\tau^+\tau^-$ and $c\bar{c}+gg$ can be measured with a precision of
$\sim 5, 10$ and $40 \%$ respectively for $M_H \sim 110$ GeV. 
For $M_H \sim 140$ GeV, BR$(H \ra WW^*)$ becomes sizeable and can be
experimentally determined; in this case the absolute magnitude of the
$b$ coupling can be derived since the $HWW$ coupling is fixed by the
production cross section. 
The Higgs coupling to top quarks, which is the largest coupling in
the SM is directly accessible in the process $\ee \ra
t\bar{t}H$. For $M_H \lsim 130$ GeV, $\lambda_t$ can be
measured with a precision of about 10 to 20\% at $\sqrt{s}\sim 500$ GeV
with $\int {\cal L} \sim 50$ fb$^{-1}$. For $M_H \gsim 350$ GeV, the
$Ht \bar{t}$ coupling can be derived by measuring the $H \ra t\bar{t}$
branching ratio at higher energies. 

Finally, the measurement of the trilinear Higgs self--coupling,
which is the first non--trivial test of the Higgs potential, is
accessible in the double Higgs production processes $\ee \ra ZHH$ and
$\ee \ra \nu \bar{\nu}HH$. However, the cross sections are rather
small and very high luminosities [and very high energies in the second
process] are needed. 

\subsubsection*{6.2 Studies at the LHC $[9]$}

In the ``low mass" range, the Higgs boson will appear as a very narrow
bump in the $\gamma \gamma$ invariant mass spectrum. The ATLAS and CMS 
collaborations claim a $\gamma \gamma$ invariant mass resolution of 
approximately 1 GeV; so the Higgs boson mass will be measured with a
good accuracy if it is detected via its two--photon decay mode. \s

For masses above 250 GeV, the Higgs boson width will be greater than 
the experimental $4l^\pm$ resolution and can therefore be measured directly. 
This allows the determination of the $HWW$ and $HZZ$ couplings [assuming 
that are related by SU(2) custodial symmetry] since the $H \ra t\bar{t}$
branching ratio is rather small. Since the $4l^\pm$ rate is proportional
to $\sigma(gg \ra H) \times$ BR($H \ra ZZ)$, one could then determine
$\Gamma (H \ra gg)$ which allows to extract the $Ht\bar{t}$ coupling
[the $Hgg$ couplings is dominantly mediated by the top quark loop 
contribution]. Some ratios of couplings could also be determined by
considering the processes $gg\ra H, qq \ra WH$ and $gg \ra t\bar{t}H$
with the subsequent decays $H \ra \gamma \gamma$ or $b\bar{b}$. 
 
\subsection*{7. Summary}

At the hadron collider LHC, the Standard Model Higgs boson can, in
principle, be discovered up to masses of ${\cal O}(1$~TeV). While the
region $M_H \gsim 130$ GeV can be easily probed through the $H \ra 4l^\pm$
channel, the $M_H \lsim 130$ GeV region is difficult to explore and a
dedicated detector as well as a high--luminosity is required to isolate
the $H\ra \gamma \gamma$ decay. SUSY Higgs bosons are more difficult to
search for, especially if decays into SUSY particles are kinematically allowed.
In this case, it is possible that no Higgs particle will be found in some 
areas of the MSSM parameter space. \s

$\ee$ linear colliders with energies in the range of $\sim 500$ GeV are
ideal instruments to search for Higgs particles in the mass range below
$\sim 250$ GeV. The search for the Standard Model Higgs particle can be
carried out in several channels and the clean environment of the
colliders allows to investigate thoroughly its properties. In the MSSM,
at least the lightest neutral Higgs particle must be discovered and the
heavy neutral and charged Higgs particles can be observed if their
masses are smaller than the beam energy. Once the Higgs bosons are
found, the clean environment of $\ee$ colliders allows to study at
great details the fundamental properties of these particles. In this
respect, even if Higgs particles are found at LHC, high energy $\ee$ 
colliders will provide an important information which make them 
complementary to hadron machines. 

\bigskip 

\nn {\bf Acknowledgements}: \s

\nn We thank the organizers of this school, in particular Marek Jezabek, 
for their warm hospitality and for the very nice and friendly atmosphere 
of the meeting.

\end{document}